\documentclass[alpha-refs]{wiley-article}

\pdfoutput=1

\usepackage{siunitx}
\usepackage{subfigure}

\usepackage{fixltx2e}
\usepackage{subfigure}

\newcolumntype{+}{!{\vrule width 2pt}}

\newlength\savedwidth

\newcommand\thickhline{\noalign{\global\savedwidth\arrayrulewidth\global\arrayrulewidth 2pt}%
\hline
\noalign{\global\arrayrulewidth\savedwidth}}

\title{The ETS challenges: a machine learning approach to the evaluation of simulated financial time series for improving generation processes}

\author[1]{Javier Franco-Pedroso}
\author[1]{Joaquin Gonzalez-Rodriguez}
\author[2]{Maria Planas}
\author[2]{Jorge Cubero}
\author[2]{Rafael Cobo}
\author[2]{Fernando Pablos}

\affil[1]{Audias Research Group, Escuela Politécnica Superior, Universidad Autónoma de Madrid, Madrid, 28049, Spain}
\affil[2]{ETS Asset Management Factory, Pozuelo de Alarcón, Madrid, 28223, Spain}

\corraddress{Javier Franco-Pedroso, Madrid, Spain}
\corremail{javierfrancopedroso@gmail.com}

\fundinginfo{ETS Asset Management Factory, Project Name: ``Tecnologías de tratamiento de señal para mercados financieros''}

\runningauthor{Javier Franco-Pedroso et al.}

\begin{document}

\maketitle

\begin{abstract}

This paper presents an evaluation framework that attempts to quantify the ``degree of realism'' of simulated financial time series, whatever the simulation method could be, with the aim of discover unknown characteristics that are not being properly reproduced by such methods in order to improve them. For that purpose, the evaluation framework is posed as a machine learning problem in which some given time series examples have to be classified as simulated or real financial time series. The ``challenge'' is proposed as an open competition, similar to those published at the Kaggle platform, in which participants must send their classification results along with a description of the features and the classifiers used. The results of these ``challenges'' have revealed some interesting properties of financial data, and have lead to substantial improvements in our simulation methods under research, some of which will be described in this work.\par

\keywords{Machine Learning; Empirical properties; Multivariate time series; Financial engineering; assets simulations}
\end{abstract}

\section{Introduction}

Simulation methods are widely used in financial applications for several purposes; for example, if asset prices are modeled as stochastic processes of some given type, confidence intervals around them can be estimated at a given future time. The goodness of those data generating processes is usually evaluated by checking if their posterior distribution fit that of real data or they do not. While this is a necessary condition, it is not sufficient, as other properties related to the time evolution of the time series (for example, auto-correlation of returns or absolute returns) are not being evaluated. Moreover, there may be some other properties of financial time series that could be completely unknown, and thus we do not know where to look in order to check if simulated series behave as real ones.\par

In this paper we present an approach that goes the other way around: instead of check if some known properties of real financial data are observed in the simulated ones, we tackle the problem of whether some simulated series can be distinguished or not from real ones. This can be addressed as a binary classification problem in which we have to distinguished between two classes, namely real and simulated financial time series. Thus, machine learning techniques can be applied to automatically extract features and train classifiers on large datasets, with the aim of properly modeling the differences between both classes based on these features. In this way, the process of comparing among different generating processes becomes completely factual, and many features (possibly some unknown as being discriminant) can be used. More over, if one specific system obtain significant good classification results, we can then look at the features and their relation with the generating process in order to improve the model or the simulation method, or maybe learn some interesting property of real financial time series unknown until now.\par

In Section \ref{sec:challenges_overview} we further motivate the need of such an evaluation framework and introduce its characteristics. Two editions of the ``challenges'' have been carried out until now, presenting differences in both asset types used (investment funds or stocks) and generation methods tested. Both are described in this work, in Sections \ref{sec:first_challenge} and \ref{sec:second_challenge}, respectively, as the aim of this paper is to highlight the lessons learned from each one. Then, the features and classifiers that obtained better results in the challenges are used to compare different generation methods in Section \ref{sec:methods_comparison} as an objective evaluation framework to test to what extent their simulated series behaves as real ones. Finally, conclusions and future work are summarized in Section \ref{sec:conlcusions}.\par

\section{ETS challenges overview and general evaluation framework}
\label{sec:challenges_overview}

The ETS challenges arise as a consequence of the need for the evaluation of simulation methods for financial time series in a blind and factual way. The most common way to check the goodness of simulation methods, from an empirical point of view, is to check if some set of known properties of real financial time series (as, for instance, the well-known ``stylized facts'' [\cite{cont01,cont07}], [\cite{chakraborti07}]) are present on the simulated ones, but this way of proceed presents several problems.\par

First, it is hard to define how to properly measure the property we are trying to quantify. For example, a well known stylized fact is that the distribution of returns presents heavier tails than a Gaussian one [\cite{mandelbrot63}], and this is quantified by the kurtosis statistic (fourth moment of the distribution). However, there exist an issue when computing this and other statistics, as a sufficient number of samples must be available in order to properly estimate it, as the statistic may not converge for the given sample size. Moreover, the degree of deviation from the normal distribution depends on the time interval between return values [\cite{mandelbrot63, mandelbrot97}], so the same property should be checked at different time scales, what may need different sample sizes in order to the statistic converges, and this is needed to be checked for several properties. This involve to individually compare many metrics between real and simulated time series, and to make a decision on whether the differences observed are acceptable or not, usually by means of some statistical hypothesis test. However, as the final goal of our simulations is to obtain simulated time series that behave as real ones, and not to test the fitness of a model in order to explain the time series behavior, the comparison can be better summarized by deciding whether real and simulated time series are distinguishable or they are not.\par

On the other hand, and even more important, if we just check the stylized facts we are constraining the search of differences to a set of already known properties, while there may be some other unknown important properties that we are just ignoring because we have not observed them previously in real time series. If instead of looking for what we already know, we just look for differences between simulated and real time series, we may found some interesting property or behavior shared by real financial data. With this aim, the goal of checking the goodness of a simulation method is tackled through an open competition posed as a binary classification problem in which a set of examples, consisting in raw return values, have to be classified as real or simulated financial time series.\par

For every challenge, two balanced sets of real and simulated time series are given to participants: one of them  (the training set) is provided along with the true class labels for developing purposes, while the other one is unlabeled (the testing set). For this latter set, participants should run their feature extractors and classifiers, developed with the aid of the training set, and provide a score for every time series segment indicating the probability of the segment to belong to one of the two classes. Answers must be submitted within a month beginning at the challenge release date, including a description of the feature extractors and classifiers used. Classification results for every submitted system are evaluated by means of the Area Under the Curve (AUC) of their Receiver Operating Characteristic (ROC) [\cite{auc_roc}]. For this metric, values close to 0.5 mean that outputs of the classifier are almost random, while values close to 1 mean accuracy almost perfect (values close to 0 also indicate almost perfect discriminative properties, but outputs pointing to the opposite class).\par

\begin{figure}[h!]
\begin{center}
\begin{minipage}{\textwidth}
\begin{center}
\includegraphics[width=10cm]{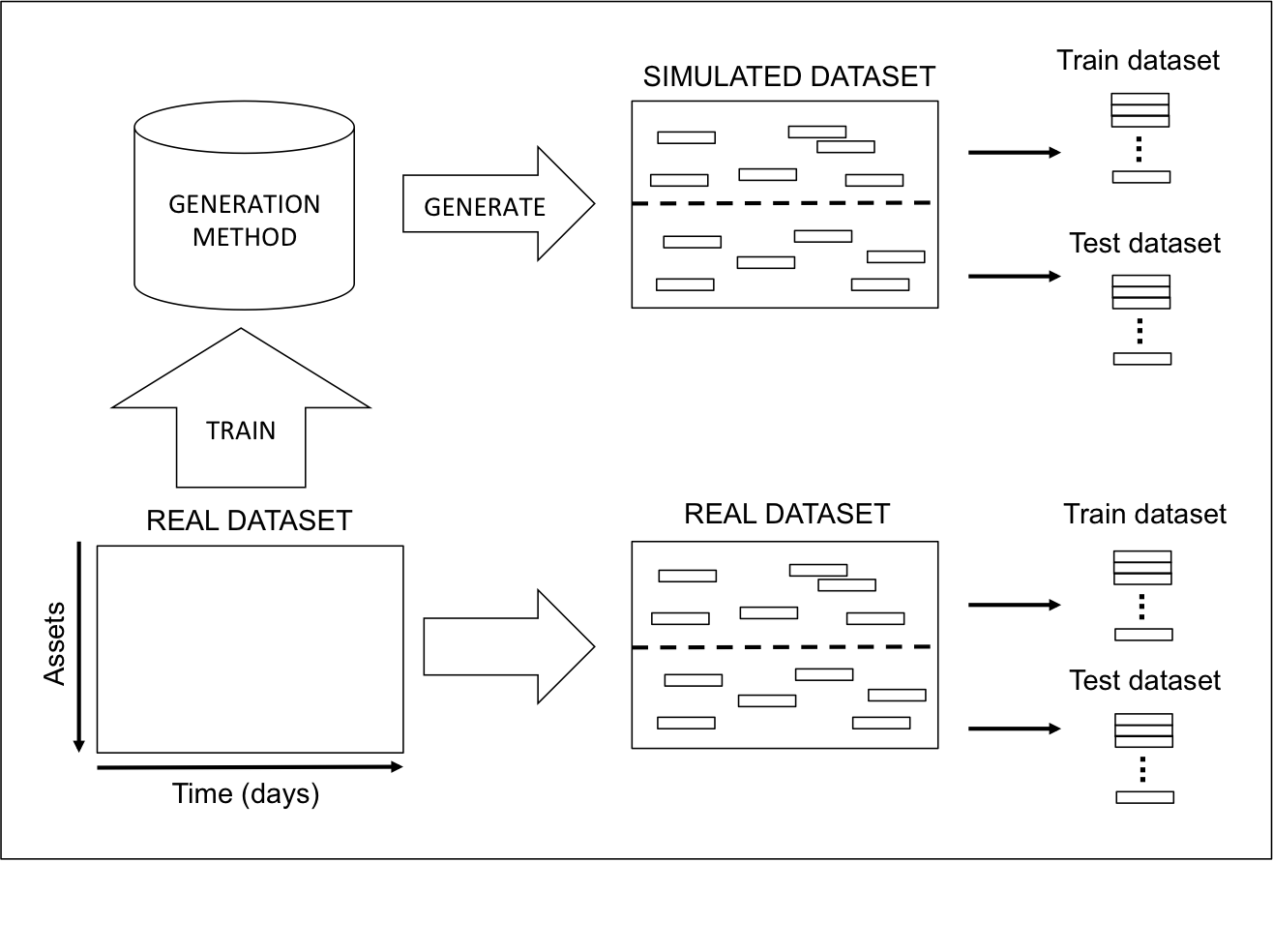}
\end{center}
\end{minipage}
\caption{Data partitioning scheme.
\label{fig:data_partitioning}}
\end{center}
\end{figure}

Both training and testing datasets comprises 6000 time series segments of 260 returns per class, as shown in Table \ref{tab:datasets}. Time series segments are extracted at random from a larger dataset, either real or simulated, but training and testing datasets are independent for both classes, as the segments for each purpose (train or test) are extracted from different time series (different investment funds or different stocks). However, they may share the time period and may come from the same market (same type of investment fund or same index). Generation methods are trained on the whole real dataset used for the challenge (both training and testing subsets), and simulations are generated in the same proportion (same number of simulations per investment fund or stock). However, the generation methods tested may not be fitted to a particular time series but to a set of them, so there may not be a one-to-one correspondence between real and simulated time series. Figure \ref{fig:data_partitioning} shows how training and test subsets are generated.\par

\begin{table}[!ht]
\centering
\caption{Composition of provided datasets.}
\begin{tabular}{|c|c|c|c|}
\thickhline
\bf{Subset} & \bf{\# examples/class} & \bf{Segments length} & \bf{Class labels} \\ \thickhline
Train & 6000 & 260 & Provided \\ \hline
Test & 6000 & 260 & Not provided \\ \hline
\end{tabular}
\label{tab:datasets}
\end{table}

\newpage

\section{2016 Challenge: detection in the context of investment funds}
\label{sec:first_challenge}

\begin{figure}[h!]
\begin{center}
\begin{minipage}{\textwidth}
\begin{center}
\subfigure[Fixed-income funds.]{
\resizebox*{7cm}{!}{\includegraphics{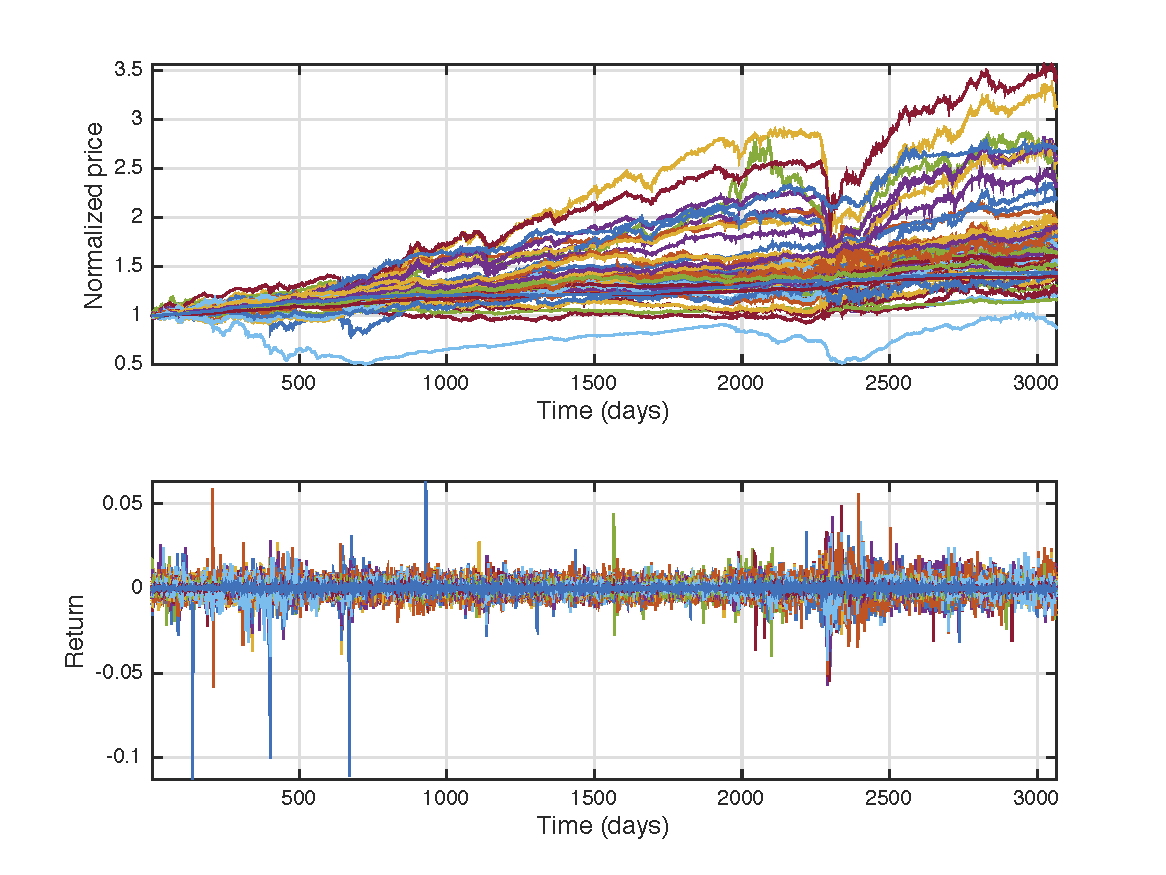}}\label{fig:1st_period}}
\subfigure[Equity funds.]{
\resizebox*{7cm}{!}{\includegraphics{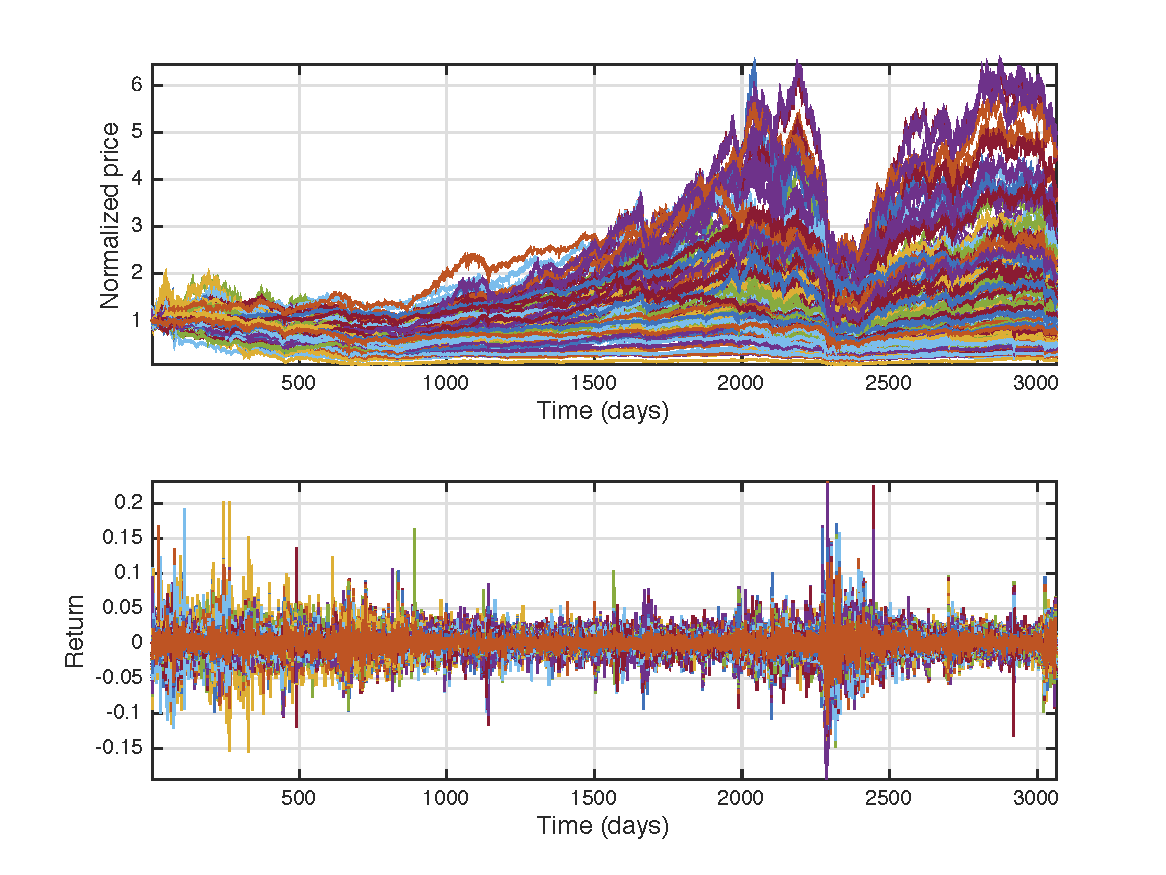}}\label{fig:3rd_period}}
\caption{Fixed-income and equity funds used in the 2016 ETS Challenge.
\label{fig:funds_datasets}}
\end{center}
\end{minipage}
\end{center}
\end{figure}

The first edition of the ETS Challenges was focused on generation methods for investment funds. Time series from two types of investment funds were used: fixed income and equity funds. Real time series sets used for the challenge are illustrated in Figure \ref{fig:funds_datasets}, showing the time series of both prices (upper panel) and returns (lower panel). For a better visualization, prices have been forced to start at a price value $p(t = 0) = 1$.\par

Fixed income subset involve 64 different funds, while the equity subset involve 198 different funds. The dataset build by merging time-aligned series from both types of investment funds was used first to train the generation process summarized in the next subsection, and then split to obtain the training and testing datasets as described in the previous section.\par

\subsection{Tested simulation method}

The simulation method used in the first edition of the ETS Challenges was an earlier version of that described in [\cite{franco18}]. The generation process can be summarized as follows:\par

\begin{itemize}
\item{\textbf{Analysis stage}: the whole multivariate training data set (each dimension being a different time series) is split into several time periods based on the trend changes estimated ex-post over the averaged time series (equally-weighted market index). Then, for each trend, a non-overlapping sliding window is used to compute mean vectors and covariance matrices from the multivariate returns (again, each dimension being a different time series) within each window. This sequence of $N_{w}$ mean vectors and covariance matrices (being $N_{w}$ the number of windows) constitutes the ``model'' of the trend.}
\item{\textbf{Synthesis stage}: first, a random sequence of alternating trends (upwards and downwards), among those obtained in the analysis stage, is hypothesized. Then, for each trend, random multivariate returns are generated by drawing multivariate samples from Gaussian distributions whose parameters are updated according to the sequence of windows observed in the analysis stage.}
\item{\textbf{New assets generation stage}: by following the two previous stages, simulated versions of the original dataset can be generated, keeping the correlations between the given time series thanks to the covariance matrices. In order to generate additional artificial assets with similar correlation properties, a PCA-based procedure is used. PCA is performed first in order to decompose the original set of time series, $R$, into eigenvectors (transformation matrix, $W$) and components (projected time series, $R'$). Then, the transformation matrix is enlarged by adding artificial eigenvectors generated from a multivariate gaussian distribution with mean and covariance  matrix obtained from $W$, leading to a new transformation matrix $W'$ with as much eigenvectors as the desired number of time series in the final simulated dataset. Finally, the components $R'$ are projected back into the original space to obtain the simulated dataset with the desired dimensions.}
\end{itemize}

Examples of simulated datasets obtained with this generation method, and an exhaustive analysis of their empirical properties can be found in [\cite{franco18}] for stock time series. From now on, this will be referred as the \textit{method 1} in order to distinguish it from the generation process followed in the second edition of the ETS Challenges (described in Section \ref{sec:second_challenge}).\par

\subsection{Submitted systems and results}

For this edition of the ETS Challenges, only a few systems where submitted, most of them not being able to distinguish between real and simulated time series, and accompanied with shallow descriptions of the development process followed by participants. For those systems, no further analysis was done. However, one of the submitted systems achieved a very high performance in our classification task (0.95 AUC), which was further analyzed in order to relate the features used with some possible shortcomings of the generation process.\par

Hopefully, this submission was deeply described by the participant, who performed in the training set an analysis of the discrimination capability provided by the features used. These features consist of 100 coefficients of the auto-correlation function (ACF) of each sample (260 return values). Each sample was subtracted from a longer one, so in this case, the ACF was a relevant characteristic in a local context. The participant found that the first principal components of each class (real and simulated) significantly differed in the training set, being this a fact that allow to distinguish between them using an ensemble of 40 k-nearest neighbor (kNN) classifiers [\cite{knn}], with $k=1$, based on the cosine distance between features' samples.\par

As mentioned in [\cite{franco18}], the \textit{method 1} does not follow an auto-regressive approach to reproduce the time series behavior, since the most reported empirical property regarding this statistic is, in fact, that they do not present significant auto-correlation as its value quickly decays for the first time lags [\cite{pagan96}]. This was also observed in [\cite{franco18}] for the simulated times series. However, it seems that, while the ACF presents insignificant values for both types of time series (real and simulated), there is still a difference in how these values behave in simulated time series compared to real ones, which allow to easily distinguish between them.\par

\subsection{Post-evaluation analysis}
\label{ssec:lessons_2016}

In order to corroborate the discriminant capabilities of the best-performing submitted system, several experiments were performed on a different time series set, consisting of stocks from the S\&P 500 index (this dataset was the one used in the second edition of the ETS Challenges, and is described in Section \ref{sec:second_challenge}). It was observed that the system was still able to distinguish between real and simulated samples with high accuracy, when a protocol similar to that one used in the challenge was followed. In order to discard any possible error or bias in the random extraction process by which samples from both classes were extracted, several experiments were performed involving only real time series. By doing this, we were trying to confirm if the features used were capturing some property shared by real financial data or, conversely, a specific particularity of the samples extracted and used with different purposes (train or test). If the features were capturing such a general property, there should not be any partition of real data that, considered as different classes, could be classified with high accuracy (that is, a classifier should not be able to distinguish between them). For that purpose, three different experiments were performed:\par

\begin{figure}[h!]
\begin{center}
\begin{minipage}{\textwidth}
\begin{center}
\subfigure[\textbf{Experiment 1}: examples for training and testing purposes come from the same time series subset.]{
\resizebox*{6.5cm}{!}{\includegraphics{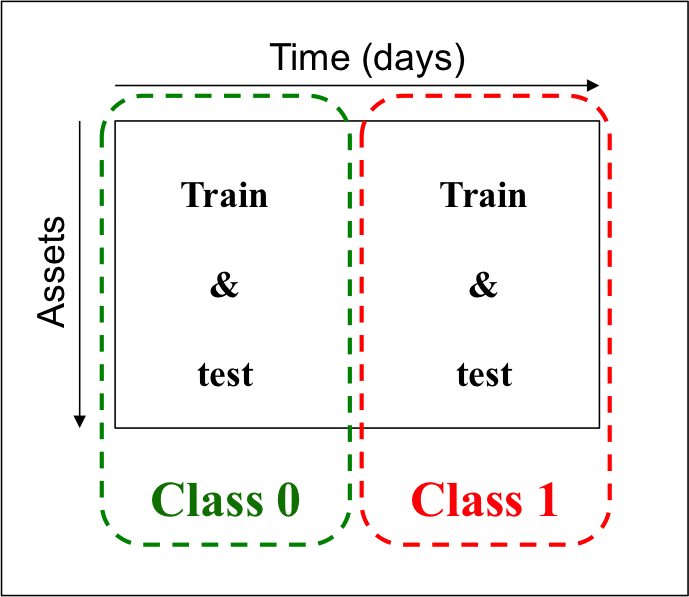}}\label{fig:exp1}}
\subfigure[\textbf{Experiment 2}: examples for training and testing purposes come from different time series subsets.]{
\resizebox*{6.5cm}{!}{\includegraphics{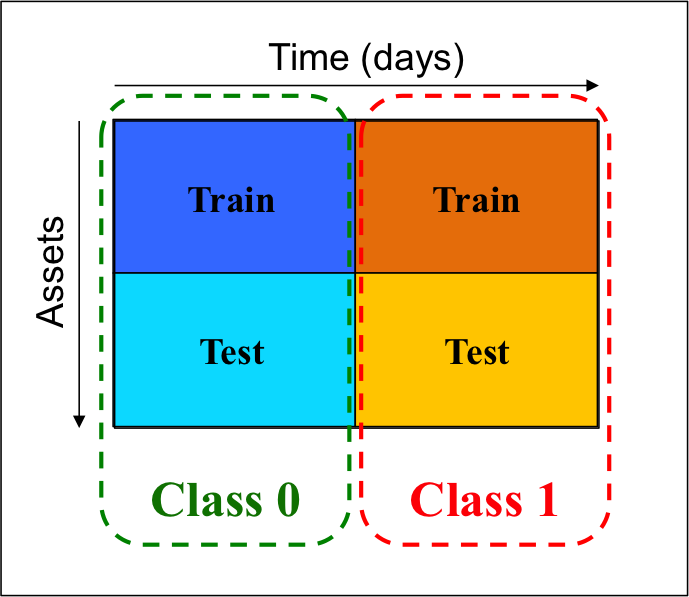}}\label{fig:exp2}}
\caption{Two classification experiments, involving only real data, in which classes represent different time periods.
\label{fig:exp1_and_2}}
\end{center}
\end{minipage}
\end{center}
\end{figure}

\begin{figure}[h!]
\begin{center}
\begin{minipage}{\textwidth}
\begin{center}
\subfigure[\textbf{Experiment 3}: examples for training and testing purposes come from the same time period.]{
\resizebox*{7cm}{!}{\includegraphics{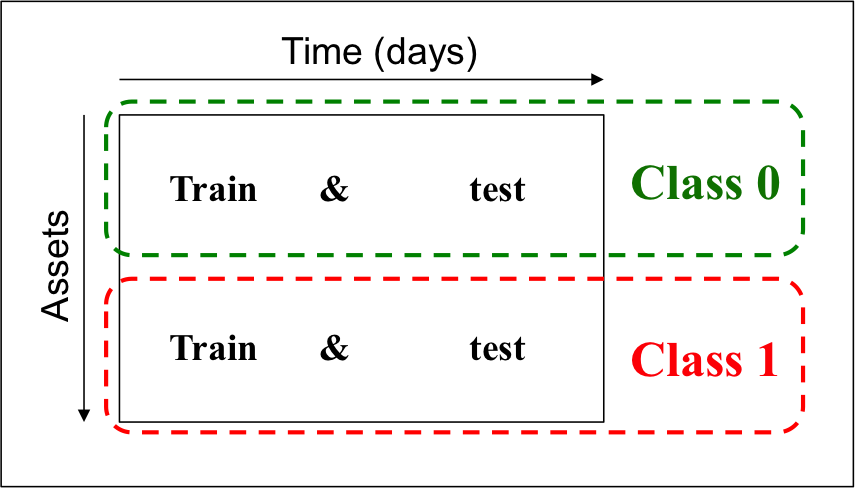}}\label{fig:exp3}}
\subfigure[\textbf{Experiment 4}: examples for training and testing purposes come from different time periods.]{
\resizebox*{7cm}{!}{\includegraphics{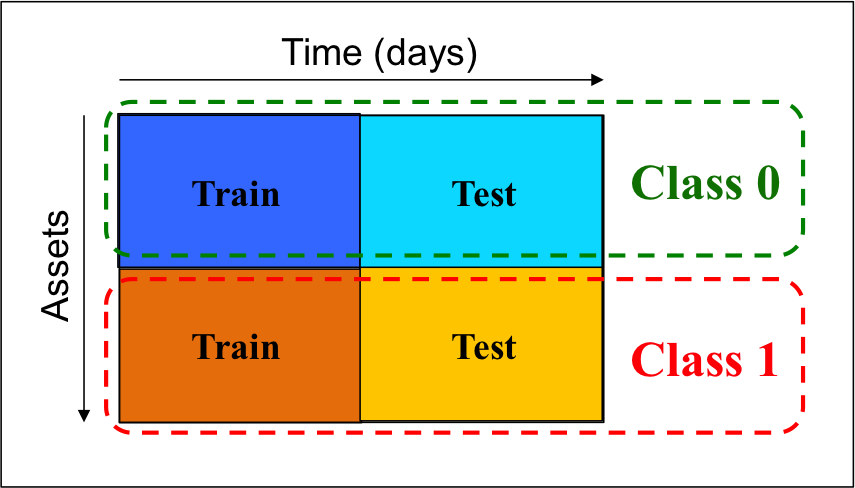}}\label{fig:exp4}}
\caption{Two classification experiments, involving only real data, in which classes represent different time series subsets.
\label{fig:exp3_and_4}}
\end{center}
\end{minipage}
\end{center}
\end{figure}

\begin{itemize}
\item{\textbf{Experiment 1}: the whole dataset was divided into two different time periods, being each period assigned to a different class. Examples for both training and testing purposes were extracted from the same subset (see Figure \ref{fig:exp1}).}
\item{\textbf{Experiment 2}: the whole dataset was divided into two different time periods, being each period assigned to a different class. For each class, data were further divided into two different time series subsets for training and testing purposes (Figure \ref{fig:exp2}).}
\item{\textbf{Experiment 3}: the whole dataset was divided into two different time series subsets, being each subset assigned to a different class. Examples for both training and testing purposes were extracted from the same subset (Figure \ref{fig:exp3}).}
\item{\textbf{Experiment 4}: the whole dataset was divided into two different time series subsets, being each subset assigned to a different class. For each class, data were further divided into two different time periods for training and testing purposes (Figure \ref{fig:exp4}).}
\end{itemize}

Results for these experiments are shown in Table \ref{tab:experiments_2016}. As it can be seen, it is possible to divide the real dataset in a way that data from different subsets can be distinguished, by classifying them as belonging to different classes, indicating that the features used do not capture a general property of real time series but rather particular differences between specific subsets. These differences notoriously arise when samples for different classes are extracted from different time periods (experiments 1 and 2), even though time series are shared among different classes. However, it is much more difficult (experiment 3) to distinguish between classes when the time period is shared, and even impossible (experiment 4) when training and testing subsets for each class come from the same time period (that is, they are overlapped among classes, but they do not between train and test within each class).\par 

\begin{table}
\centering
\caption{Experiments performed involving only real time series.}
\begin{tabular}{|c|c|c|}
\thickhline
\bf{Experiment} & \bf{Classes represent} & \bf{AUC} \\ \thickhline
1 & Different time periods & 0.95 \\ \hline
2 & Different time periods (train/test) & 0.90 \\ \hline
3 & Different assets & 0.75 \\ \hline
4 & Different assets (train/test) & 0.48 \\ \hline
\end{tabular}
\label{tab:experiments_2016}
\end{table}

The fact that the auto-correlation is similar for different time series if they are close in time could be partially explained by the usual presence of cross-asset correlations between different assets of the same type or coming from the same market [\cite{plerou99}], as they evolve over time in a similar way. However, similar experiments performed on simulated datasets by using the \textit{method 1} showed that any partition made on the dataset did not provide subsets that could be distinguished or classified as belonging to different classes, even though cross-asset correlations were properly reproduced [\cite{franco18}]. The reason for such auto-correlation pattern not being reproduced on simulated data is that the \textit{method 1} do not follow an autoregressive approach but only attempt to match distributional properties. In order to reproduce such a behavior of real time series, a new generation method was developed, which is described in the next Section.\par

Returning to the ability of the submitted system to distinguish between real and simulated datasets, it was observed that the classifier also achieved a high performance (0.9 AUC) if ACF coefficients were computed for absolute return values instead, revelaing that significant differences in volatility clustering [\cite{mandelbrot63}], which is an already well known ``stylized fact'' [\cite{cont01,cont07}], [\cite{chakraborti07}], could be found as well. Thus, both systems (the submitted system and this latter one) were used as our \textit{reference} systems (Reference 1 and 2, respectively), or sanity checks, for every generation process developed from that moment.\par

\newpage

\section{2017 Challenge: detection in the context of stocks}
\label{sec:second_challenge}

\begin{figure}[h!]
\begin{center}
\begin{minipage}{\textwidth}
\begin{center}
\includegraphics[width=10cm]{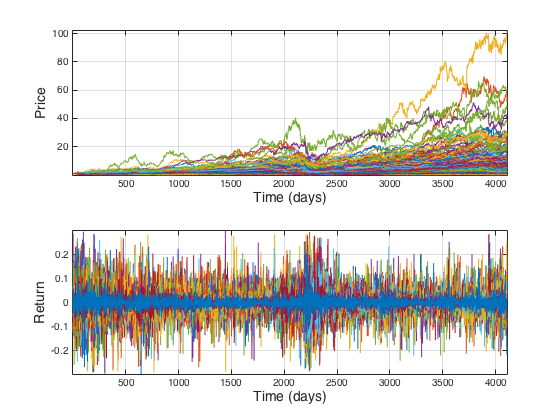}
\caption{Stocks dataset used in the 2017 ETS Challenge.
\label{fig:stocks}}
\end{center}
\end{minipage}
\end{center}
\end{figure}

The second edition of the ETS Challenges was focused on testing our generation methods on stock data. Particularly, the main dataset used consist of the daily prices/returns between 01/01/2000 and 04/29/2016 of a set of 330 stocks that have been part of the S\&P500 index at some time within this given period. This dataset is illustrated in Figure \ref{fig:stocks}, showing the time series of both prices (upper panel) and returns (lower panel). For a better visualization, stock prices have been forced to start at a price value $p(t = 0) = 1$.\par

The whole dataset was used as the training dataset for the generation methods in order to obtain the simulated dataset. Then, it was split into two halves (different stocks, same period, as indicated in Figure \ref{fig:data_partitioning}) to extract time series segments for both train and test datasets used in the challenge. Only segments coming from this dataset were included as training data, while testing data also included time series segments from different datasets (considered as out-of-set examples):\par

\begin{itemize}
\item same stocks as those in the main dataset but coming from a different time period.
\item stocks form a different market (EUROSTOXX index), same or different time period from the main dataset.
\end{itemize}

\subsection{Tested simulation method}
\label{sec:method_2017}

In order to overcome the main issue that our previous generation process presented (\textit{Method 1}), as revealed in the previous edition of the ETS Challenges (Section \ref{sec:first_challenge}), a different approach was followed. Similarly to the previous approach, the generation method can be summarized into the following stages:\par

\begin{figure}[h!]
\begin{center}
\begin{minipage}{\textwidth}
\begin{center}
\subfigure[Multivariate vectors in \textit{Method 1}.]{
\resizebox*{6.5cm}{!}{\includegraphics{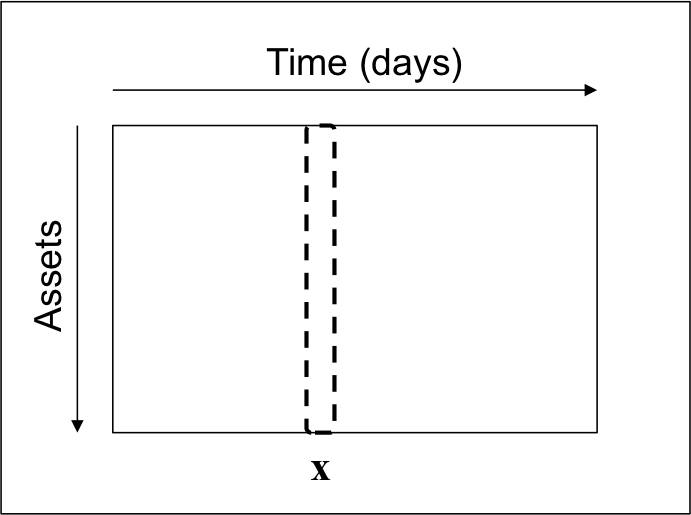}}\label{fig:method1}}
\subfigure[Multivariate vectors in \textit{Method 2}.]{
\resizebox*{6.5cm}{!}{\includegraphics{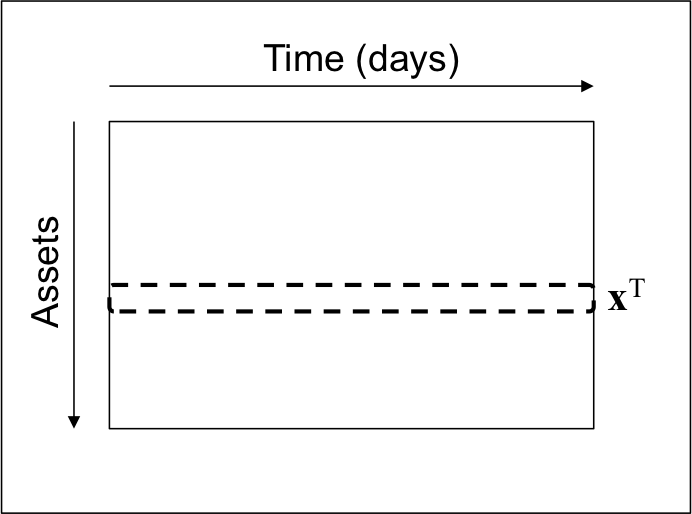}}\label{fig:method2}}
\caption{Comparison of the information modeled by multivariate vectors ($\mathbf{x}$) in \textit{Method 1} (a) and in \textit{Method 2} (b). For \textit{Method 2}, the represented dataset is assumed to come from an isolated trend.
\label{fig:methods_comparative}}
\end{center}
\end{minipage}
\end{center}
\end{figure}

\begin{itemize}
\item{\textbf{Analysis stage}: the whole multivariate training data set is split into several time periods based on the trend changes estimated ex-post over the averaged time series (equally-weighted market index), as it was done in \textit{Method 1}. However, the multivariate data within a trend is processed in a very different way. Instead of considering data as a time sequence of multivariate return values (each dimension being a different asset), each time series (or asset) within a trend is considered a multivariate sample itself in which the return values at different time steps are seen as different dimensions (see Figure \ref{fig:methods_comparative}). Then, a mean vector and covariance matrix (whose dimensions depends on the length of the trend) are obtained for each trend, what constitutes the ``model'' of the trend, and represents the average behavior of the market within this time period. In this way, the average auto-correlation is captured by the covariance between dimensions (time steps).}
\item{\textbf{Synthesis stage}: as it was done for our previous approach, a random sequence of alternating trends (upwards and downwards) is hypothesized first. Then, for each trend, random return values are generated by drawing multivariate samples from a Gaussian distribution with mean vector and covariance matrix equal to that observed in the analysis stage. Note that, in this approach: \textit{i}) the whole trend for a specific asset is generated at once by drawing a multivariate sample, and \textit{ii}) there is no need for a procedure such as the PCA-based one used by \textit{Method 1} if more assets want to be generated, as we can simply generate more multivariate samples for the same trend.}
\end{itemize}

While this approach does not explicitly model the correlation between assets, it has been observed that the simulated data do present cross-asset correlation, as different assets are different samples drawn from the same multivariate Gaussian distribution and thus the time evolution within a trend is similar for different time series. On the other hand, the time series produced by following the previously described stages do not show a key feature of financial time series as the heavy tails are. For this reason, the following additional steps were included:\par

\begin{figure}[h!]
\begin{center}
\begin{minipage}{\textwidth}
\begin{center}
\subfigure[Step 1: real CDF estimation.]{
\resizebox*{6.2cm}{!}{\includegraphics{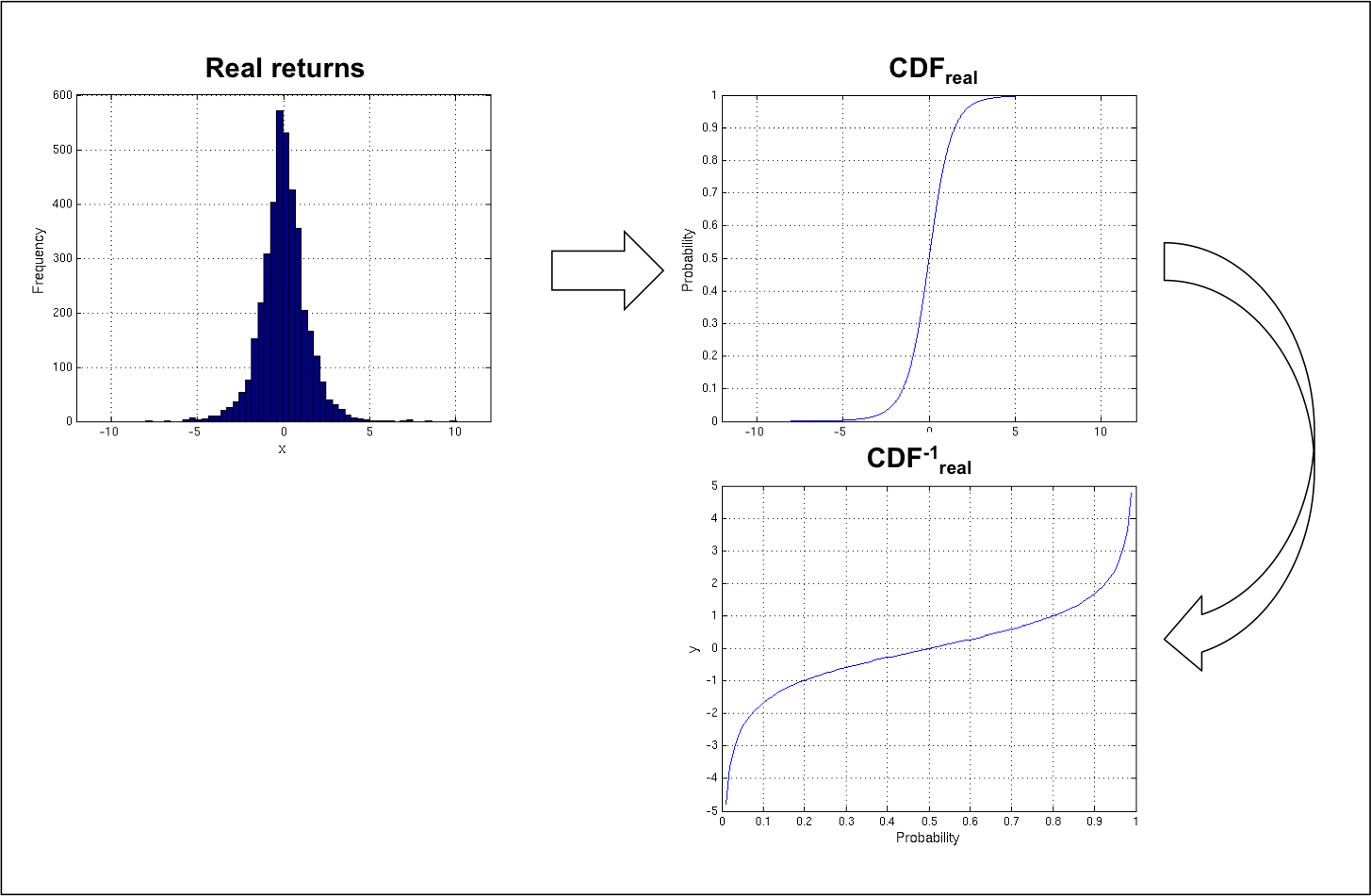}}\label{fig:step_1}}
\subfigure[Step 2: histogram equalization for simulated return values.]{
\resizebox*{5.7cm}{!}{\includegraphics{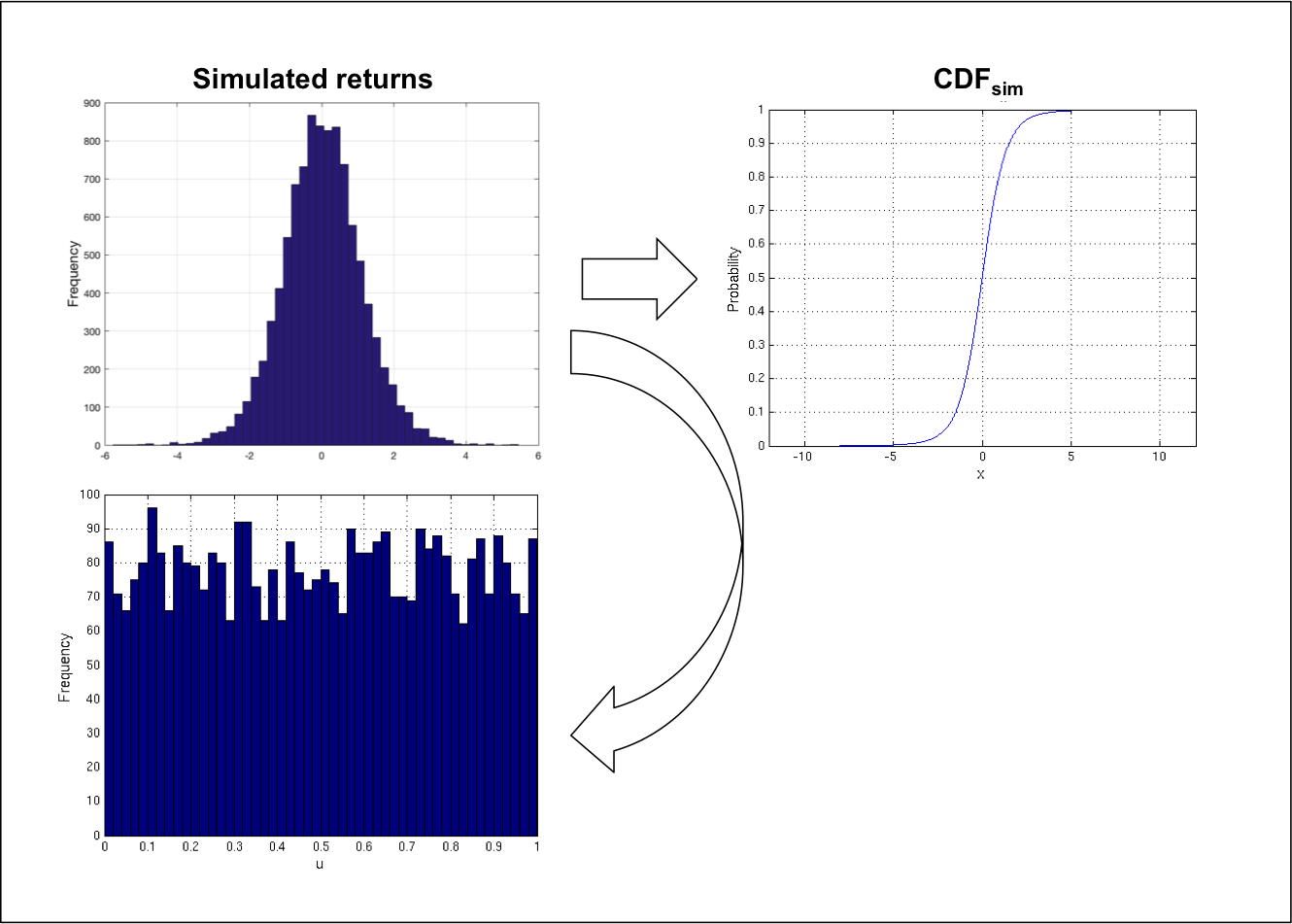}}\label{fig:step_2}}
\subfigure[Step 3: transformation of equalized histogram to fit real CDF.]{
\resizebox*{6cm}{!}{\includegraphics{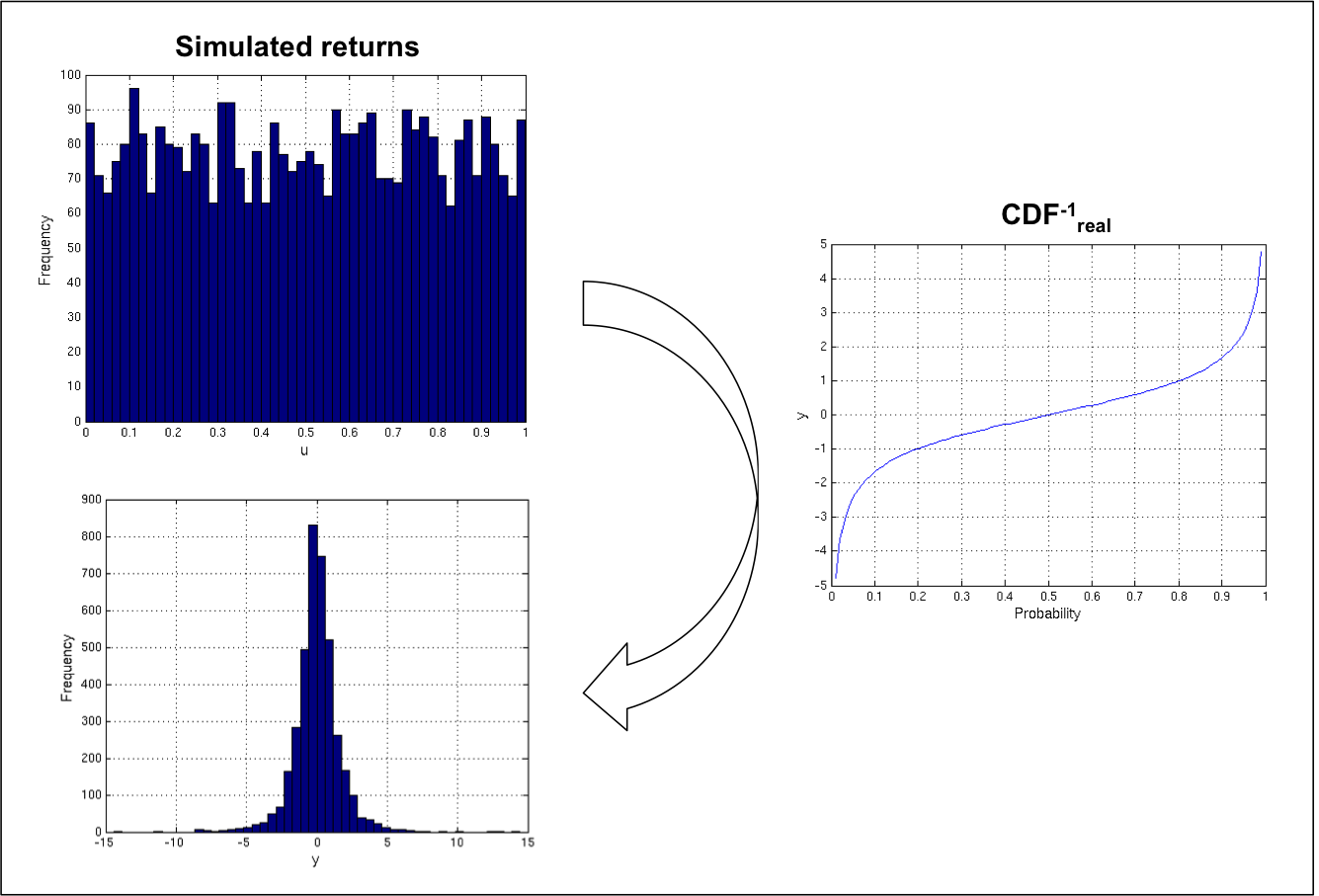}}\label{fig:step_3}}
\caption{Additional steps performed to fit real CDF on simulated return values. Bar figures represent histograms, while line figures represent estimated CDF or its inverse, for a specific return time series within a trend.
\label{fig:histogram_eq}}
\end{center}
\end{minipage}
\end{center}
\end{figure}

\begin{itemize}
\item{In the analysis stage, the cumulative distribution function (CDF) of the time series within each trend are estimated (Figure \ref{fig:step_1}).}
\item{In the synthesis stage, after the generation of random samples from the multivariate Gaussian distribution, the histogram of the return values from each sample (one time series within a trend) is first equalized (Figure \ref{fig:step_2}), and finally transformed by applying the inverse of the CDF estimated in the synthesis stage (Figure \ref{fig:step_3}).}
\end{itemize}

In this way, the simulated asset returns within a trend better fits the distribution of real time series. However, this fitness is not ``complete'', as revealed by our reference system developed in order to check distributional properties (``Reference system 3'' in Table \ref{tab:2017_reference_results}), among others. This system uses as features several statistics (average, standard deviation, median, kurtosis, skewness), the Hurst exponent [\cite{mandelbrot69}], the Sharpe ratio [\cite{sharpe66}] and some other metrics that also quantifies the shape of the distribution (percentage of return values bigger or smaller than some specific thresholds); the classification technique for this system consist of 100 bootstrapp-aggregated decision trees (\textit{bagged trees}) [\cite{Breiman2001}]. On the other hand, as it can be seen for the ``Reference systems'' 1 and 2, the problem previously observed regarding both the auto-correlation of asset returns and volatility clustering was completely solved by the approach followed in \textit{Method 2}.\par

\begin{table}[!ht]
\centering
\caption{Summary of reference systems and results.}
\begin{tabular}{|c|c|c|c|}
\thickhline
\bf{Reference system} & \bf{Features (\#)} & \bf{Classifier} & \bf{AUC} \\ \thickhline
1 & 100 ACF coeffs. returns (100) & Ensemble of 40 KNNs & 0.57\\ \hline
2 & 100 ACF coeffs. absolute returns (100) & Ensemble of 40 KNNs & 0.55\\ \hline
3 & Statistics, Hurst exponent and Sharpe ratio (8) & 100 bagged trees &  \bf{0.76} \\ \hline
\end{tabular}
\label{tab:2017_reference_results}
\end{table}

\subsection{Submitted systems and results}

Seven submissions were received in this edition of the ETS Challenge, one of them including two different systems. Those systems are describe in the following paragraphs, including both the features and the classifiers used.\par

\begin{itemize}

\item{\textbf{System 1:}
\begin{itemize}
\item Features. It used 4 normality tests (Person's chi-squared [\cite{pearson}], Anderson-Darling [\cite{anderson1952}], Lilliefors [\cite{Lilliefors67}] and sftram), maximum drawdown (MDD) [\cite{mdd}], stationarity test and volatility clustering.
\item Classifier. It used a gradient boost classifier [\cite{gradient_boost_1, gradient_boost_2}].
\end{itemize}}

\item{\textbf{System 2:}
\begin{itemize}
\item Features. It used five sets of features: products and differences between consecutive return values ($r(t)$ and $r(t+1)$) in time series, as well as the rolling standard deviation within small windows (2 and 3 return values) and the return values themselves. Each of these sets were finally sorted in ascend order. Feature vectors were of dimension 1295.
\item Classifier. It used a gradient boost classifier.
\end{itemize}}

\item{\textbf{System 3:}
\begin{itemize}
\item Features. It used the amplitude frequency values extracted through a 260-point Fast Fourier Transform (FFT) [\cite{fft}], and retains only half of the spectrum. Feature vectors were of dimension 130.
\item Classifier. It used a 130-component Gaussian Mixture Model (GMM) [\cite{mixture_dist}] to model each class, and obtained the score for each class as the probability density of the test segment for that class. Finally, returned the ratio for the positive class.
\end{itemize}}

\item{\textbf{System 4:}
\begin{itemize}
\item Features. It used amplitude frequency values extracted through a 260-point FFT, and retaining only half of the spectrum. Feature vectors were of dimension 130.
\item Classifier. It used a one hidden layer neural network (NN) [\cite{Bishop_NNs}] with 65 rectified linear units (ReLU) [\cite{relu}], and one output unit with sigmoid activation. The network was trained with mean squared error (MSE) as the cost function.
\end{itemize}}

\item{\textbf{System 5:}
\begin{itemize}
\item Features. It used several statistic metrics (mean, volatility, skewness, kurtosis) and a random hypothesis test.
\item Classifier. It used a binary regression tree.
\end{itemize}}

\item{\textbf{System 6:}
\begin{itemize}
\item Features. It used the difference between the autocorrelation (ACF) and the partial-autocorrelation (PACF) for the first 10 coefficients of each segment. Feature vectors were of dimension 10.
\item Classifier. It obtained a probability score for the segment belonging to the positive class as the ratio of coefficients classified as belonging to that class, applying an heuristic threshold.
\end{itemize}}

\item{\textbf{System 7:}
\begin{itemize}
\item Features. It used the \textit{tsfresh} [\cite{tsfresh}] library (open-source feature extractor for time series classification) to extract 222 features.
\item Classifier. It used a gradient boost classifier.
\end{itemize}}

\item{\textbf{System 8:}
\begin{itemize}
\item Features. It used the amplitude frequency values extracted through a 260-point FFT. Feature vectors were of dimension 260.
\item Classifier. It used k-Nearest Neighbors (kNN) classifier.
\end{itemize}}

\end{itemize}

Table \ref{tab:2017_results} summarizes the features and the classifier used for each system along with the results obtained in the challenge. It is interesting to note that only those systems that used gradient boost classifiers (Systems 1, 2 and 7) were able to distinguish between real and simulated time series, while the rest of the submitted systems provide almost random outputs ($\sim0.5$ AUC). Among those best-performing systems, there was, however, a significant gap in performance (AUC) between System 1 (0.61) and Systems 2 (0.82) and 7 (0.89) that can be partially explained by the number of features used: just 7 features in System 1 versus hundreds in System 7 or even more than a thousand in System 2. However, most of the features used by System 1 consist of normality tests, a property which is already known that real time series do not present, and that has been avoided in simulated time series as explained in the previous section. The importance of the classifier used is also revealed by the results obtained by System 5, which uses similar features to those used by our third reference system described in the previous section but obtains much worse performance (0.49 vs. 0.76 AUC). On the other hand, some features as the FFT values seem to not provide discriminant capabilities, as the results obtained for a variety of classifiers (GMMs in System 3, NN in System 4 and kNN in System 8) are almost random. Finally, System 6 uses similar features to those used by our first reference system described in the previous section, but a lower number of them, and it uses a much simpler classifier, leading to a very low performance.\par

\begin{table}[!ht]
\centering
\caption{Summary of submitted systems and results.}
\begin{tabular}{|c|c|c|c|}
\thickhline
\bf{System} & \bf{Features (\#)} & \bf{Classifier} & \bf{AUC} \\ \thickhline
1 & Normality tests, MDD, stationarity test, vol. clust. (7) & Gradient boost & 0.61 \\ \hline
2 & Sorted differences, products, and rolling STD (1295) & Gradient boost & \bf{0.82} \\ \hline
3 & FFT absolute values, half spectrum (130) & GMM &  0.49 \\ \hline
4 & FFT absolute values, half spectrum (130) & NN &  0.49 \\ \hline
5 & Stats and random hyp. test (5) & Binary regression tree &  0.49\\ \hline
6 & PACF minus ACF coefs. (10) & Heuristic threshold &  0.5\\ \hline
7 & Max, min, stats, value counts, ... (222) & Gradient boost &  \bf{0.89}\\ \hline
8 & FFT absolute values, whole spectrum (260) & kNN & 0.51\\ \hline
\end{tabular}
\label{tab:2017_results}
\end{table}

\subsection{Post-evaluation analysis}

In order to improve the simulation method used for this Challenge, the two best-performing submitted systems  where further analyzed. Conclusions and findings obtained are exposed in this section.\par

\begin{figure}[h!]
\begin{center}
\includegraphics[width=12cm]{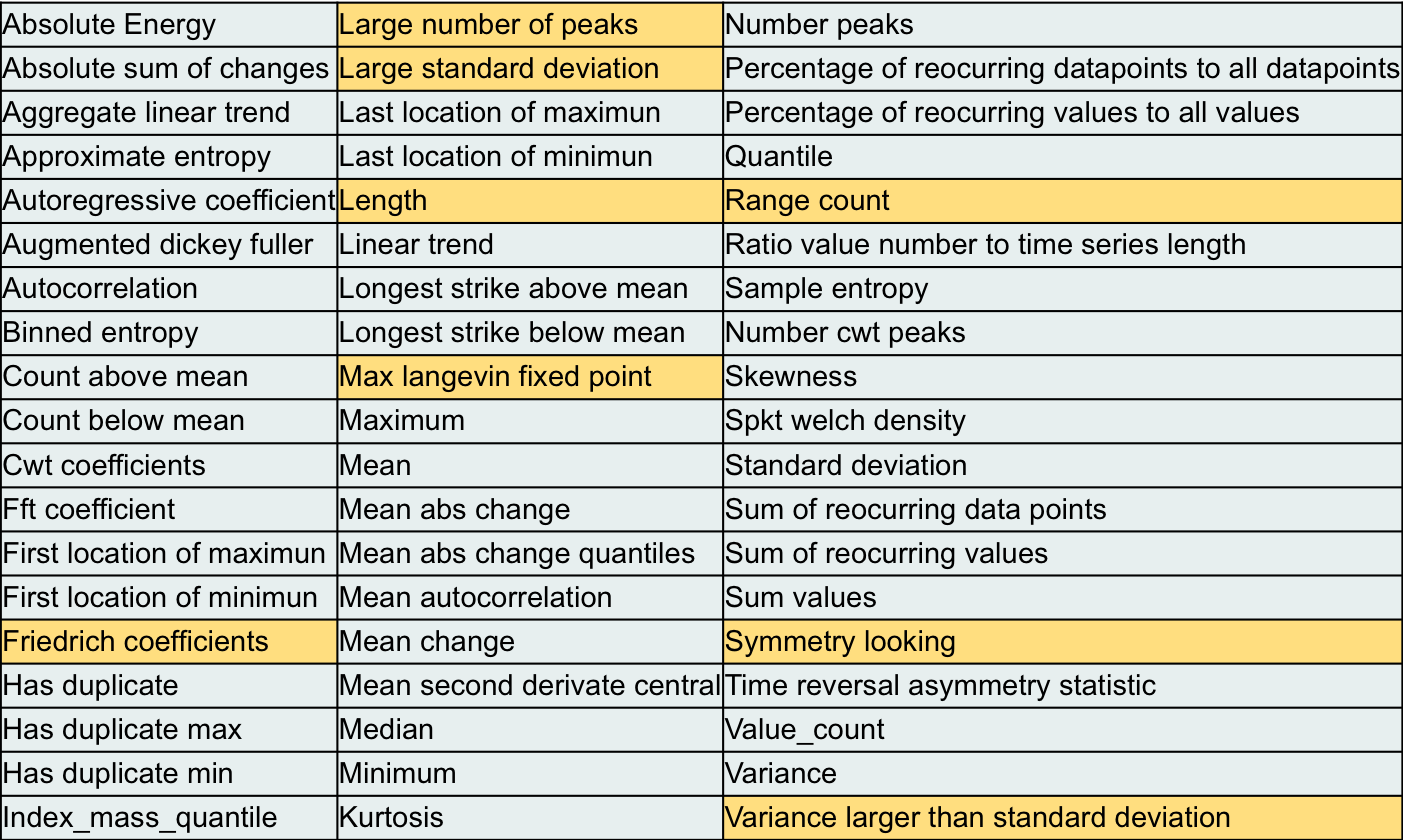}
\caption{Features removed after the data 'cleaning' process are highlighted.
\label{fig:tsfresh_features}}
\end{center}
\end{figure}

First, we look at the features used by the submitted System 7, which obtained the best performance in the Challenge. As previously mentioned, this system used an open-source library (\textit{tsfresh}) specifically designed to extract relevant features from time series [\cite{tsfresh}]. This feature extractor computes 57 sets of features (see Figure \ref{fig:tsfresh_features}), some of which are obtained for several values of the parameters they depend on, leading to a final set of 222 features. It was observed, however, that for some of those features NaN values were always obtained and converted to zero values, while some others present always the same value for every sample (for example, the feature ``length'', as all time series were equally long). Those features (highlighted in Figure \ref{fig:tsfresh_features}) were removed, without changing the performance of the system, leading to a final set of 169 feature that was further analyzed.\par

Next, the relative importance of individual features was analyzed by means of a classifier based on a bag of decision trees, which revealed that 6 predictors were significantly more important among the whole 169 set. Those features were the following:\par

\begin{itemize}
\item Percentage of reoccurring datapoints to all datapoints.
\item Sum of reoccurring datapoints.
\item Percentage of reoccurring values to all values. 
\item Sum of reoccurring values.
\item Binned entropy (max bins: 10).
\item Ratio value number to time series length (=1 if any value is repeated).
\end{itemize}

The difference between reoccurring datapoints and reoccurring values can be noticed with in following example: the time series $\{1, 4, 0.5, 1, 2.9, 1.8, 4\}$ presents 4 reoccurring datapoints (4 array positions for which the value at them also appears at another different position) but only 2 reoccurring values (2 specific values, namely the number 1 and the number 4, that also appear at other array positions).\par

All those features are related to the randomness of the return values in a time series sample: the more random the values, the more likely it is to be repeated (and the greater the entropy). As the simulation process generate return values by drawing random samples from a Gaussian distribution, and they are further transformed through a continuous function (steps 2 and 3 in Figure \ref{fig:histogram_eq}), simulated returns are assured to be completely random, not presenting any repeated values. However, the real dataset was discovered to had a high percentage of repeated values (close a 25\% of the whole training dataset), probably due to truncation in price values as they were obtained from the database. In order to ensure that this was the main difference between real and simulated data detected by System 7, the two following steps were performed:\par

\begin{figure}[h!]
\begin{center}
\includegraphics[width=10cm]{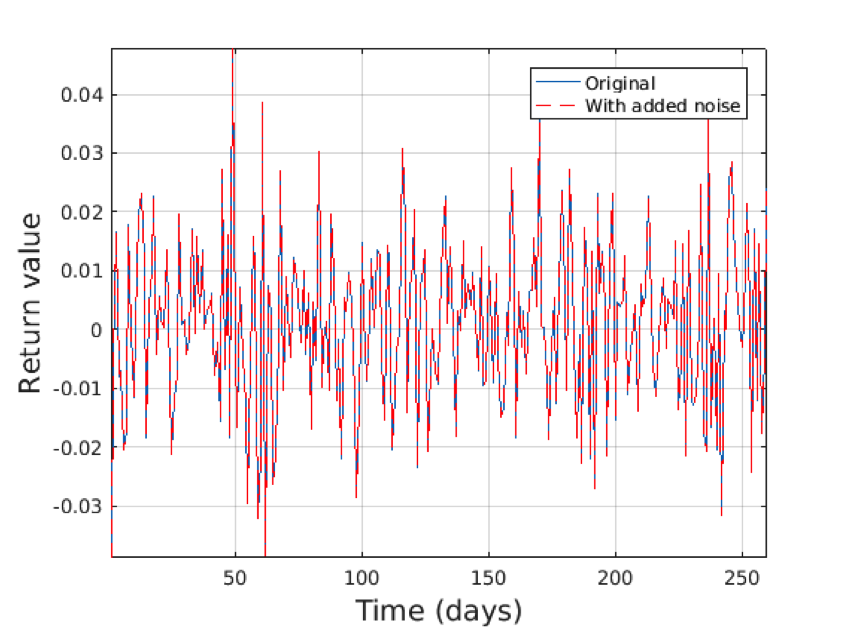}
\caption{Original time series sample and its noisy version.
\label{fig:series_with_noise}}
\end{center}
\end{figure}

\begin{itemize}
\item First, and independent feature extractor was developed to extract only the previously mentioned features, and the same classifier was used. The difference in performance between this modified system, which used only 6 features instead of 222, and the submitted one, was as low as 0.2\% AUC in absolute terms.
\item Then, a very low-power random noise was added to real time series samples to avoid repeated return values. The maximum amplitude of the noise was set to $10^{-13}$, nine orders of magnitude smaller than the minimum absolute return value in real time series ($10^{-4}$). This specific value was the minimum that lead to non repeated values in the real dataset. As it can be seen in Figure \ref{fig:series_with_noise} for a particular sample, the behavior of the time series was not changed appreciably. The same was done for simulated time series as well to assure that every input sample was processed in the same way regardless its class, which is unknown in the testing phase.
\item Finally, the system was used to process those modified time series samples replicating the Challenge protocol, obtaining a great drop in performance that lead to almost random output (0.53\% AUC), confirming therefore the initial hypothesis. On the other hand, the performance for any of the other submitted systems was not modified.
\end{itemize}

Submitted System 2 obtained the second best performance at the Challenge. As described in the previous section, the features used by this system can be grouped in the following subsets:\par

\begin{itemize}
\item Sorted return values in the time series sample (260 features).
\item Sorted products between consecutive return values (259 features).
\item Sorted differences between consecutive return values (259 features).
\item Sorted standard deviation values computed over a rolling window of length 2 (259 features).
\item Sorted standard deviation values computed over a rolling window of length 3 (258 features).
\end{itemize}

\begin{figure}[h!]
\begin{center}
\includegraphics[width=10cm]{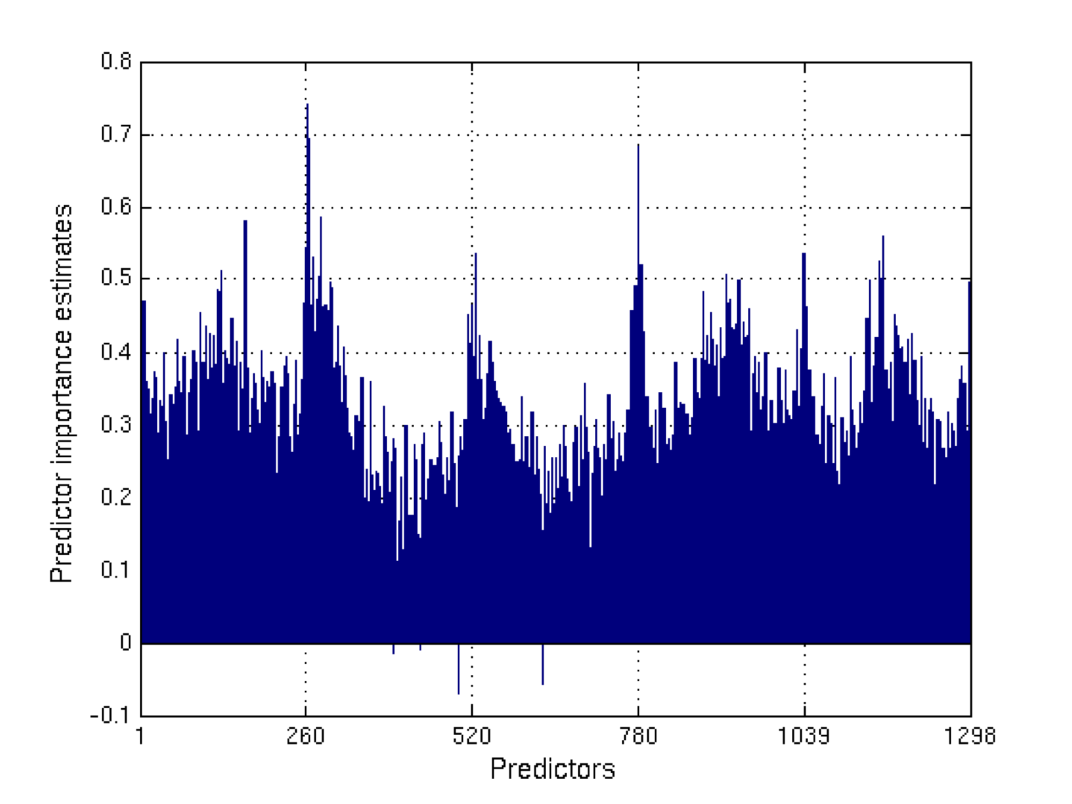}
\caption{Relative importance of individual features for submitted System 2. Predictor importance estimates represent the increase in prediction error if the values of each feature are permuted across the out-of-bag observations.
\label{fig:rel_import}}
\end{center}
\end{figure}

Similarly to the process done with submitted System 7, an analysis of the relative importance of individual features was performed first. The aim was not to look for a specific feature being particularly discriminant among the whole set, but for subsets of them. As shown in Figure \ref{fig:rel_import}, all subsets present features with higher relative importance at the edges of the feature subset, and some of them at the middle as well. It is important to note that the process of sorting feature values within each subset converts them into an approximation of the Cumulative Density Function (CDF), without normalizing counts. Then, what Figure \ref{fig:rel_import} reveals is that the distributions of those values differ between real and simulated time series, especially for extreme values of the distribution.\par

\begin{figure}[h!]
\begin{center}
\begin{minipage}{\textwidth}
\begin{center}
\subfigure[CDFs comparison for a short trend.]{
\resizebox*{5.8cm}{!}{\includegraphics{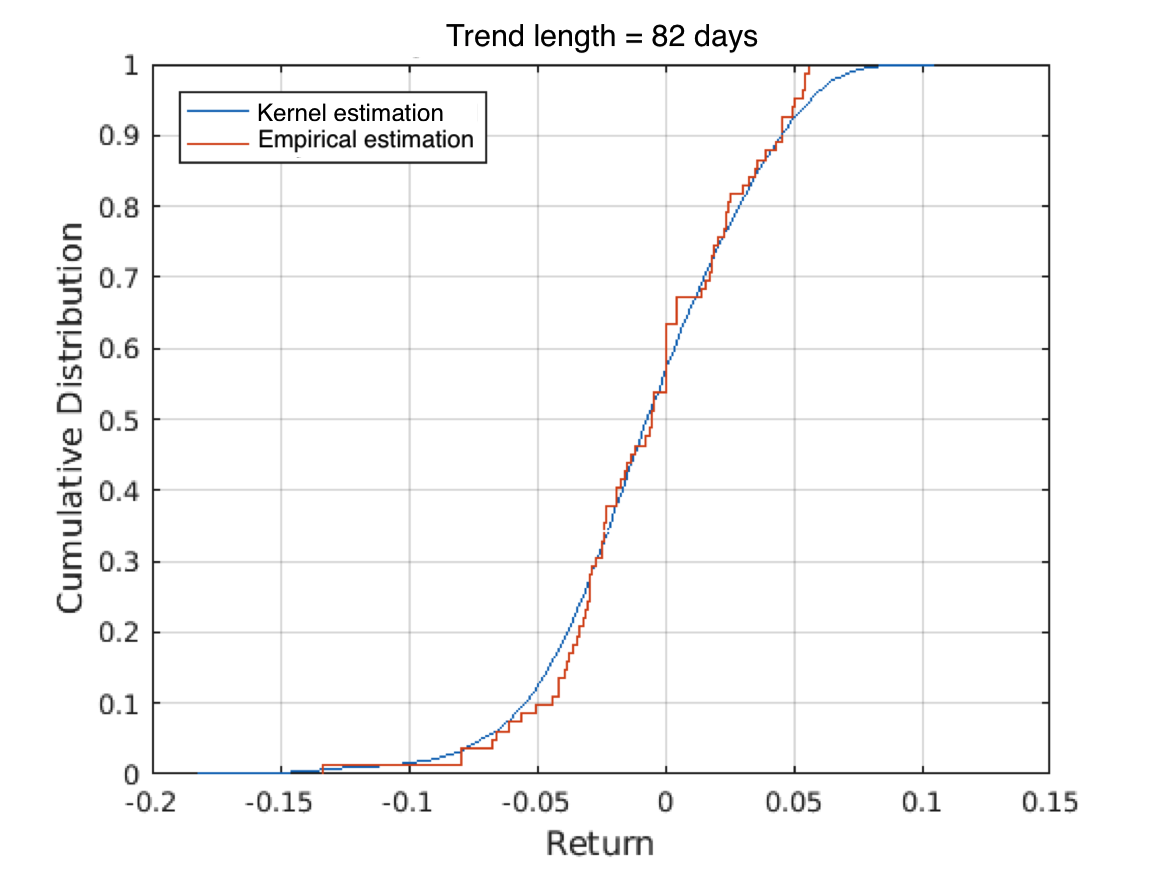}}\label{fig:diff_cdfs_1}}
\subfigure[CDFs comparison for a long trend.]{
\resizebox*{5.8cm}{!}{\includegraphics{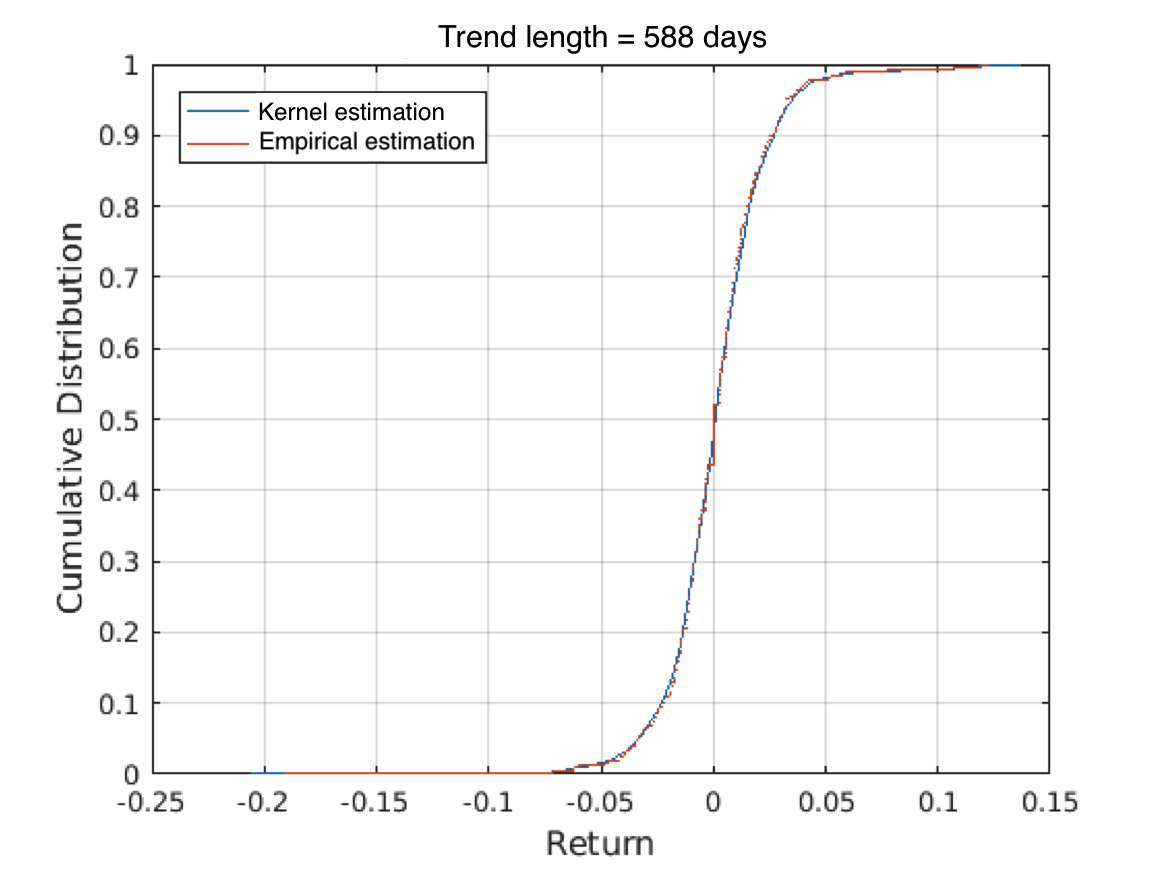}}\label{fig:diff_cdfs_2}}
\subfigure[Detail form b) around 0 return values.]{
\resizebox*{5.8cm}{!}{\includegraphics{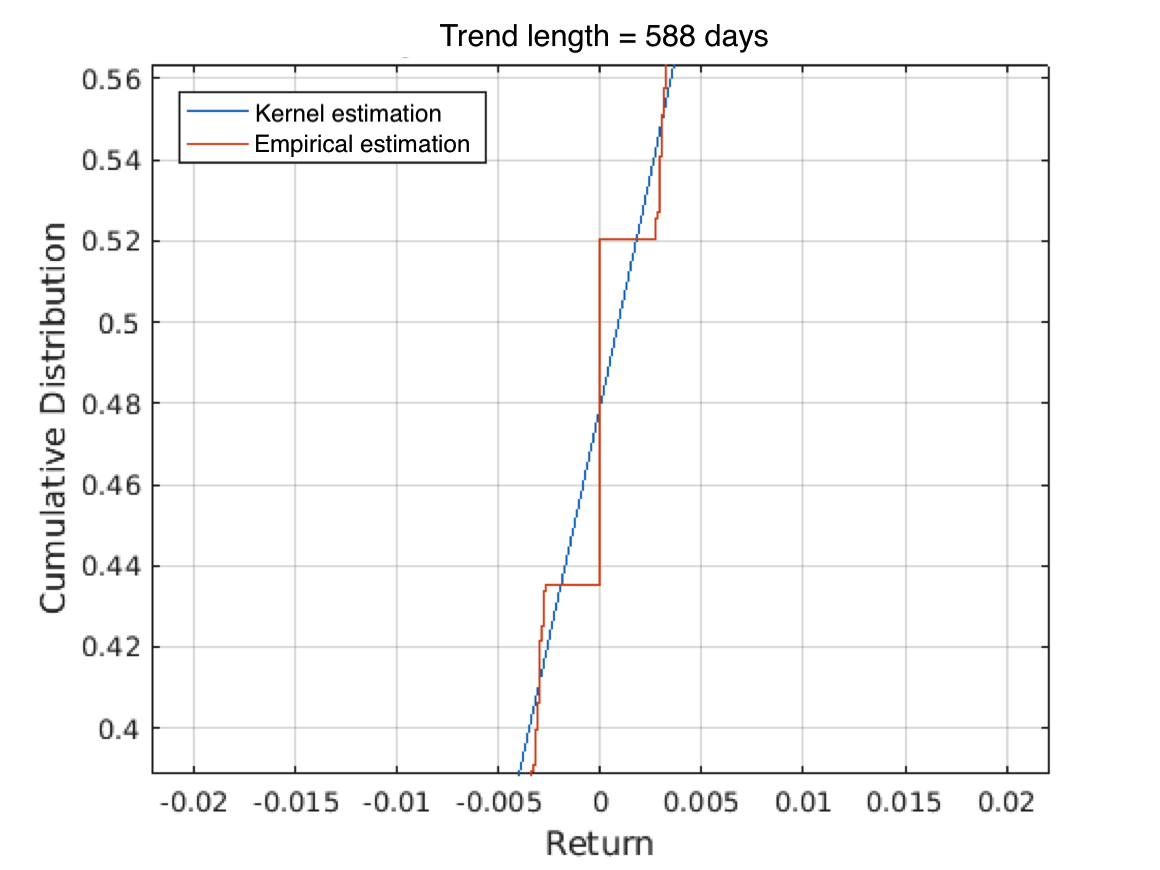}}\label{fig:diff_cdfs_3}}
\caption{Empirical and kernel estimations of the CDF for two different trends for a particular time series.
\label{fig:cdf_diffs}}
\end{center}
\end{minipage}
\end{center}
\end{figure}

With this in mind, and taking into account that the simulation process attempt to fit the distribution of return values within each trend, we looked at the differences between the ``real'' distribution (empirical CDF) and that one used to transform the simulated returns (step 3 in Figure \ref{fig:histogram_eq}), which was estimated through a kernel function (step 1 in Figure \ref{fig:histogram_eq}). Those differences are highlighted in Figure \ref{fig:cdf_diffs} for two different trends of a particular stock. As it can be seen in Figure \ref{fig:diff_cdfs_1}, high differences arise close to the extremes of the distribution for short trends, as few samples are observed and they to are concentrated around the mean value. Thus, the kernel CDF can not be properly estimated in those return ranges. When the trend is longer (Figure \ref{fig:diff_cdfs_2}), the kernel estimate is much more closer to the empirical CDF. However, if we look closer at an interval around zero, we can notice that even for long trends, differences also arise as real time series may have a large number of zero return values, as shown in Figure \ref{fig:diff_cdfs_3}. These differences explain, at least, the relative importance of predictors being at the extreme and mid-range values of the first subset of features (1 to 260 in Figure \ref{fig:rel_import}) but they could also affect the remaining ones as they are based on calculations involving consecutive return values.\par

With the aim of overcoming this issue, kernel estimates of the CDF were replaced in \textit{Method 2} by empirical estimates, leading to a better fit of the distribution of stock returns within each trend. This modification lead to a drop in performance of the submitted System 2 from 0.82 to 0.76 AUC, confirming the hypothesis that some of the features were capturing the described differences in the CDF. However, the systems still allow to distinguish between real and simulated time series. This could be due to the way auto-correlations are being modeled/reproduced by \textit{Method 2}. Many of the features are based on the relation between consecutive return values in time series. These relations are modeled in \textit{Method 2} by the covariance between adjacent dimensions. However, as the covariance matrix is computed as the sample covariance between time series segments (trends) belonging to different time series, an overall behavior is being estimated, while those relations may change from one time series to another.\par

On the other hand, it was noticed that the performance of Reference system 3 where also highly affected, dropping from 0.76 to 0.6 AUC. This is a consequence of the better fit to the CDF, as some of its features directly attempt to quantify the distribution of stock returns, especially those representing the percentage of return values smaller/larger than a given threshold (computed for several thresholds).\par

\section{Simulated-series detection-systems as an evaluation framework for generation processes}
\label{sec:methods_comparison}


As it has been shown in previous sections, the proposed challenges allow to objectively measure the goodness of a simulation method and to easily relate the features used with shortcomings in the properties modeled by the process. Similarly, we can compare among different simulation methods by choosing those systems that better performed on the proposed task. In this section, we use the Reference systems 1, 2 and 3, and the Systems 2 and 7 submitted to the second edition of the Challenges (note, however, that for System 7 we only include the most relevant features found in previous section). The methods compared, apart from our Method 2 after applying the improvements already mentioned in previous section, are the following:\par

\begin{itemize}
\item{Stochastic Differential Equation (SDE) models: Geometric Brownian Motion (GBM) and Constant Elasticity of Variance (CEV).}
\item{Univariate Generalized Autoregressive Conditional Heteroscedastic (GARCH) models [\cite{Bollerslev86}]:  exponential GARCH (EGARCH) and Glosten-Jagannathan-Runkle (GJR) model [\cite{gjr}].}
\item{Multivariate GARCH: BEKK implementation for 2 and 5 dimensions. Note that this is one of the main modern models that currently define the state of the art when generating financial series.}
\end{itemize}

Results are shown in Table \ref{tab:comparative_methods}. As the goal of the simulation methods is to generate time series as similar as possible to real ones, the better is the performance of the detection systems (AUC values), the worse is the simulation method. That is: ``good performance'' of a simulation method means low AUC value for a detection system. Then, it can be seen that \textit{Method 2} clearly outperformed the other well-known models for all the detection systems used. It is also interesting to note that results agree with previous knowledge regarding those other models: for example, SDE models do not follow an autoregressive approach, so they perform bad on Reference system 1, which is based on autocorrelation; also they do not model the time-varying behavior of variance, and so they perform bad for Reference system 2 as well. The converse happens with GARCH models, while they do not perform as well as \textit{Method 2}. GARCH models also perform better than SDE ones for Reference system 3 and submitted System 2, as they used Student's T innovations and so reproduce better statics features as heavy tails. For the multivariate implementation tested, it can be seen the effect previously reported in [\cite{franco18}]: volatility clustering is worse reproduce (see results for Reference system 2) when more dimensions are attempted to model simultaneously. For this implementation, Gaussian innovations were used, so again we have lower performance for Reference system 3.\par

\begin{table}[!ht]
\centering
\caption{AUC values for the set of systems used to detect simulated time-series. Lower performances are highlighted, indicating that simulation method generates time series more similar to real ones.}
\begin{tabular}{c|c|c|c|c|c|c|c|}
\cline{2-8}
& \multicolumn{7}{c|}{\bf AUC} \\
\cline{2-8}
& \multicolumn{2}{c|}{\bf SDE} & \multicolumn{2}{c|}{\bf GARCH} & \multicolumn{2}{c|}{\bf BEKK} & {\bf Method 2}\\
\cline{2-7}
								& GBM & CEV & EGARCH & GJR & 2-dim & 5-dim & \\ \thickhline
\multicolumn{1}{|c|}{Reference 1} & 0.81 & 0.79 & 0.61 & 0.62 & 0.62 & 0.63 & {\bf 0.55} \\ \hline
\multicolumn{1}{|c|}{Reference 2} & 0.89 & 0.89 & 0.68 & 0.66 & 0.65 & 0.76 & {\bf 0.56} \\ \hline
\multicolumn{1}{|c|}{Reference 3} & 0.98 & 0.98 & 0.83 & 0.83 & 0.96 & 0.98 & {\bf 0.6} \\ \hline
\multicolumn{1}{|c|}{System 7}    & 0.94 & 0.94 & 0.54 & {\bf 0.51} & 0.91 & 0.94 & {\bf 0.51} \\ \hline
\multicolumn{1}{|c|}{System 2}    & 0.98 & 0.98 & 0.94 & 0.94 & 0.96 & 0.98 & {\bf 0.76} \\ \hline
\end{tabular}
\label{tab:comparative_methods}
\end{table}

\section{Conclusions and future work}
\label{sec:conclusions}

This paper has presented the ETS Challenges, an open competition posed as a machine learning problem in which participants have to classify financial time series examples from a test set as real or simulated, based on models and features developed with the aid of a labeled training set. The goal of the competitions is two-fold: first, to test the goodness of our developed simulation methods in a factual manner in terms of their final goal (generate financial time series as similar to real ones as possible), and secondly, to find clues, not necessarily related with already known properties, that allow to identify their weaknesses in order to improve them.\par

In the first edition of the Challenges, one of the submitted systems showed that the simulation method used generated time series that could be easily distinguished from real ones comparing their autocorrelation coefficients in a local way. Moreover, it was observed that this system was also capable of distinguish between different subsets of real time series when they came from different time periods, revealing that, although financial time series do not present significant autocorrelation to allow predicting future price movements, they do exhibit some kind of pattern that is shared among different stocks within a given time period, and it changes over time.\par

In the second edition of the Challenges, two findings were done. On the one hand, one of the submitted systems (System 7) revealed an issue not related with the simulation process but with the real time series itself, as it detected that repeated return values were present in the original dataset. On the other hand, the submitted System 2 showed that the distributional properties of real returns, among others, were not perfectly reproduced by the simulated ones, as it was also pointed out by our Reference system 3. This allowed to focus our efforts in finding better ways of reproducing this property. Although the simulation method was improved, as shown by the drop of performance of the system, some differences remain between real and simulated time series and further research is needed.\par

Moreover, the evaluation framework defined for the challenges has allowed us to factually compare the latest version of our simulation method with some other well known and widely used methods, showing that ours performed significantly better for all of the detection systems. Finally, it can be noted that this cyclic process involving the evaluation of simulated samples and consequent improvement of the simulation method seems to be a perfect scenario for applying Generative Adversarial Networks (GANs) [\cite{goodfellow14}], which have shown very promising results for other applications, specially for image synthesis. In this way, the whole process could be fully automatized, avoiding the need for finding a good combination of features and classifier. As a counterpart, the features being more discriminative could remain hidden depending on the discriminator complexity, which would prevent us to find unknown properties of financial data.\par

\bibliography{sample}

\end{document}